\documentclass[preprint,12pt]{elsarticle}
\usepackage[dvipsnames]{xcolor}
\usepackage{xfrac}

\newcommand{\abinit}{{\textit{Ab Initio}}}

\usepackage{amssymb}
\usepackage{chemformula}

\usepackage[linktocpage=true,
 colorlinks=true, 
  pdfborder={0 0 0},
  linkcolor=blue,
  citecolor=red,
  filecolor=yellow,
  urlcolor=blue,
  bookmarks,
  pdfauthor={},
]{hyperref}

\journal{SoftwareX}

\begin{document}

\begin{frontmatter}



\title{Performing highly efficient Minima Hopping structure predictions using the Atomic Simulation Environment (ASE)}


\author[inst1]{Marco Krummenacher}
\author[inst1]{Moritz Gubler}
\author[inst1]{Jonas A Finkler}
\author[inst1]{Hannes Huber}
\author[inst1]{Martin Sommer-Jörgensen}
\author[inst1]{Stefan Goedecker}

\affiliation[inst1]{organization={Department of Physics, University of Basel},
            addressline={Klingelbergstrasse 82}, 
            city={Basel},
            postcode={4056}, 
            state={Basel},
            country={Switzerland}}

\begin{abstract}
In the dynamic field of materials science, the quest to find optimal structures with low potential energy is of great significance. Over the past two decades, the minima hopping algorithm has emerged as a successful tool in this pursuit.
We present a robust, user friendly and efficient implementation of the minima hopping algorithm as a Python library, enhancing in this way the global structure optimization simulations significantly.
Our implementation significantly accelerates the exploration the potential energy surfaces, leveraging an MPI parallelization scheme that allows for multi level parallelization.
In this scheme, multiple minima hopping processes are running simultaneously communicating their findings to a single database and, therefore, sharing information with each other about which parts of the potential energy surface have already been explored.
Also multiple features from several existing implementations such as variable cell shape molecular dynamics and combined atomic position and cell geometry optimization for bulk systems, enhanced temperature feedback and fragmentation fixing for clusters are included in this implementation.
Finally, this implementation takes advantage of the Atomic Simulation Environment (ASE) Python library allowing for high flexibility regarding the underlying energy and force evaluation.
\end{abstract}

\begin{graphicalabstract}
\includegraphics[width=0.95\textwidth]{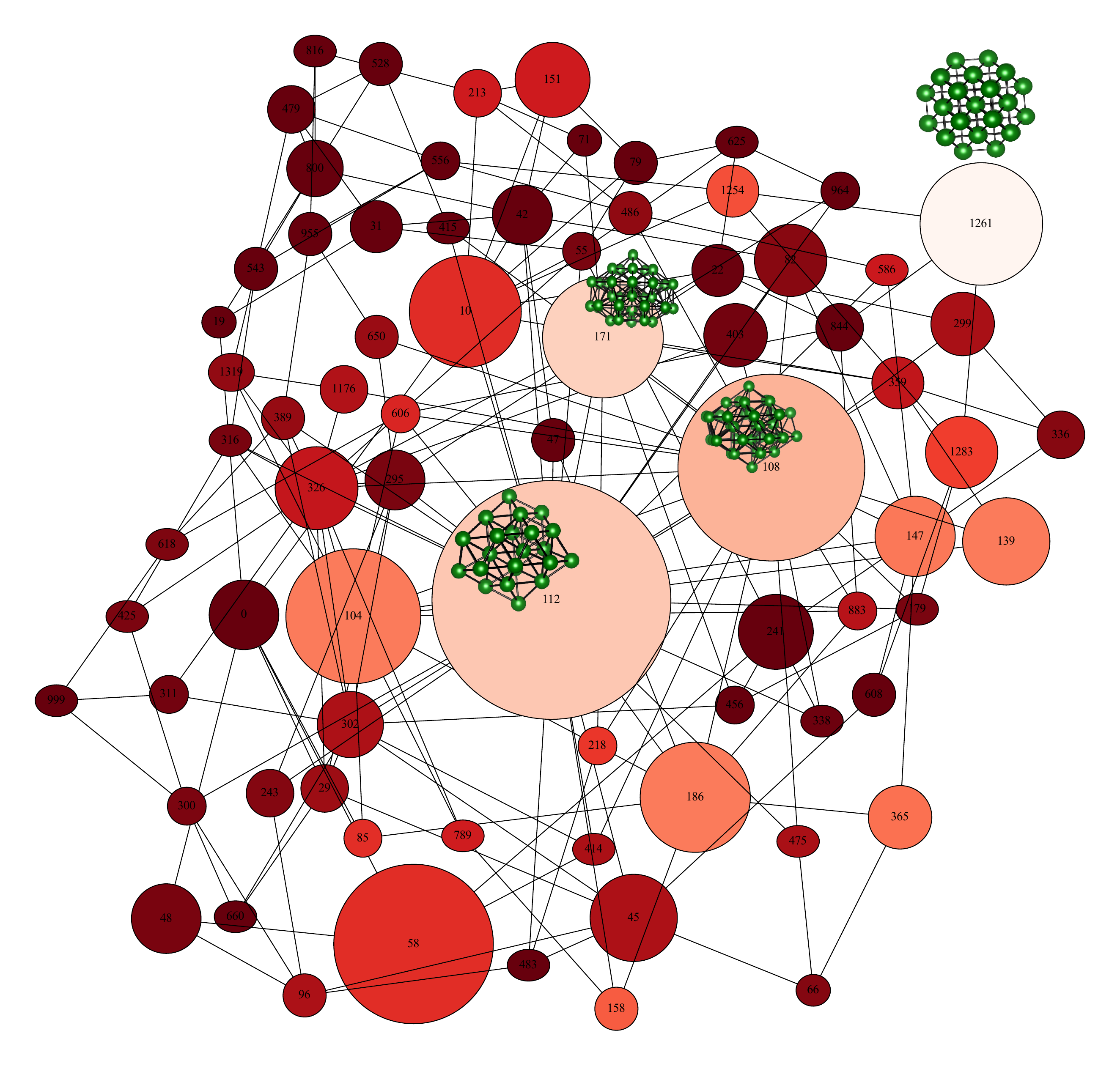} 
\end{graphicalabstract}

\begin{highlights}
\item Exploration of the potential energy surface
\item Global structure optimization
\item Reaction networks
\item High throughput
\item MPI parallelization
\end{highlights}


\end{frontmatter}


\section{Motivation and Significance}
\label{sec:Introduction}
Global structure optimization is an important task in materials science.
However, finding the structure lowest in energy can be very challenging due to the vast and high-dimensional search space and the presence of multiple local minima on a potential energy surface (PES).
For performing this task several methods have been proposed such as simulated annealing~\cite{kirkpatrick:1983}, basin hopping~\cite{wales:1997}, random structure search~\cite{pickard:2006}, meta-dynamics~\cite{martovnak:2007} and different variants of evolutionary genetic algorithms~\cite{johnston:2003, oganov:2006, bhattacharya:2013, vilhelmsen:2014, zhu:2015, curtis:2018} as implemented in the USPEX~\cite{glass:2006}, CALYPSO~\cite{wang:2012}, XtalOpt~\cite{falls:2020} and MAGUS~\cite{magus}.
Yet another method is the Minima Hopping (MH) algorithm~\cite{goedecker:2004} first proposed in 2004. 
In contrast to evolutionary algorithms MH can not get trapped in a funnel~\cite{schonborn:2009}.
When trapping is detected by the algorithm, the system will melt due to high temperatures in the MD escape part and then, once it has escaped from the present funnel, freeze into another funnel. Over the years the MH method has been applied successfully for structure prediction for a variety of cluster and bulk systems and surfaces.
By combining MH with saddle point search to determine transition structures, reaction pathways can be constructed with the so called minima hopping guided path search~\cite{schaefer:2014}. 
In many cases a biasing of the PES can be used to drive the MH algorithm faster towards a desired target structure. Thus speeding up the pathway search~\cite{de:2019, de:2020, huber:2023}.
Hence, the MH method can not only be used for global structure search but can also provide data that can be used in a post-processing step to construct reaction/transformation pathways. 
\\
In the past 20 years several versions were developed implementing a variety of features tailored specifically to the system which was studied.
Originally the algorithm was directly implemented into the density functional code BigDFT~\cite{genovese:2011} which prevents the usage of other DFT codes, machine learned potentials (MLPs) and force fields.
For a variety of DFT codes such as Quantum Espresso~\cite{giannozzi:2020}, gpaw~\cite{mortensen2005real}, cp2k~\cite{kuhne2020cp2k}, abinit~\cite{gonze2020abinit}, SIRIUS~\cite{sirius_lapw}, castep~\cite{clark2005first} as well as BigDFT and MLPs such as NequIP~\cite{batzner:20223} and Flare~\cite{vandermause:2020} an interface to the atomic simulation environment (ASE)~\cite{larsen:2017} is available.
ASE is a Python library containing a set of tools and modules for setting up, manipulating and running atomistic simulations, including Minima Hopping runs.
At present also a basic version of MH is implemented in ASE~\cite{peterson2014global}.
However, several features are implemented differently or missing such as softening of the velocities, structure distinction, fixing fragmented clusters or enhanced temperature feedback.
This motivated us to implement a user--friendly MH version which is compatible with any ASE calculator and to bring together several features implemented in different previous versions of MH.
Among those features are for example a highly efficient variable cell shape SQNM optimizer~\cite{gubler:2022} or the collection of data sets which can later be of use for the construction of MLPs.
Further, an MPI parallelization is implemented where multiple MH runs share a single database.

\section{Minima Hopping Algorithm}
\begin{figure}
    \centering
    \includegraphics[width=0.5\textwidth]{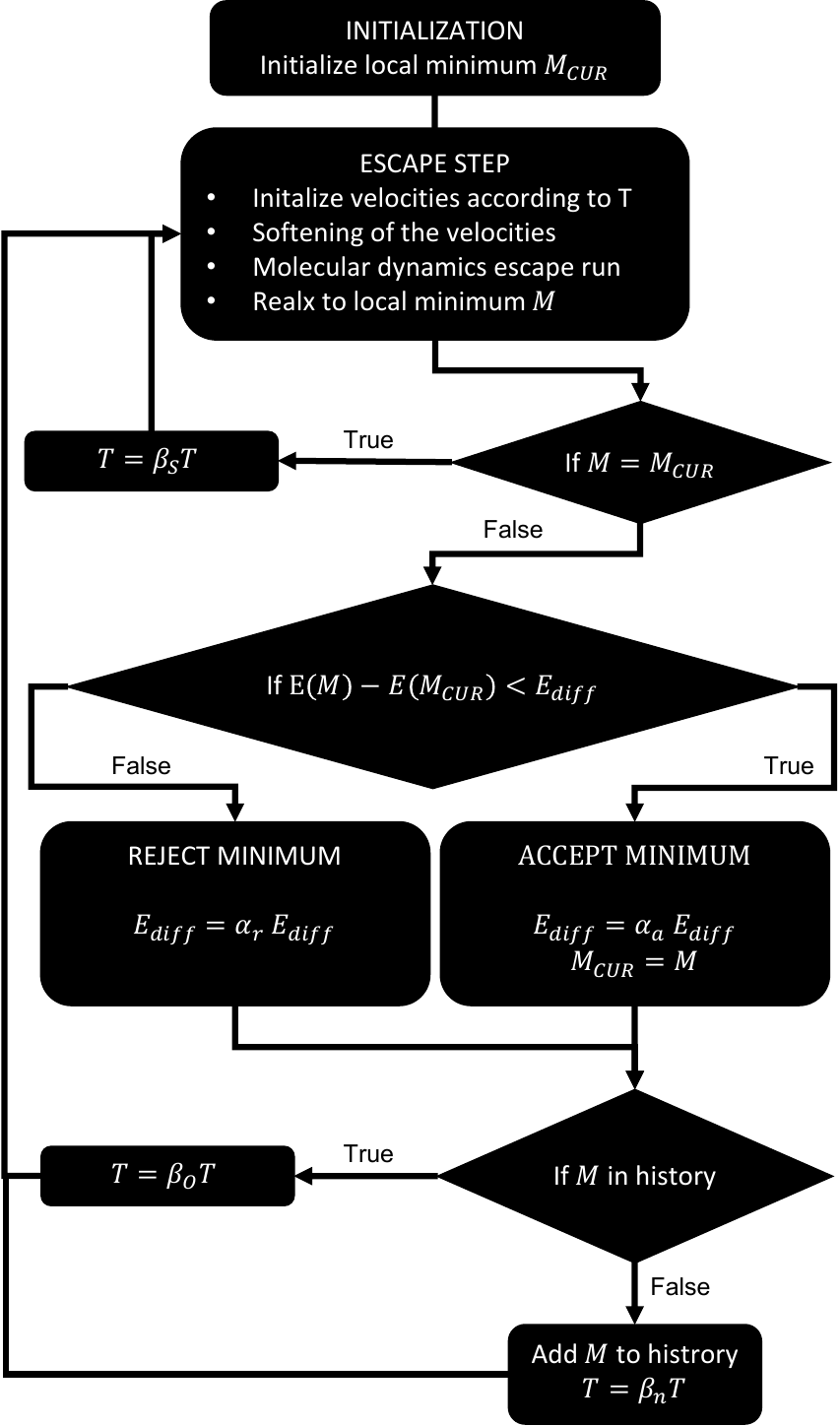}
    \caption{Minima hopping algorithm: $\alpha_{r} > 1$, $\alpha_{a}<1$, $\beta_{s} > 1$, $\beta_{o} > 1$, $\beta_{n} < 1$.}
    \label{fig:mh_algorithm}
\end{figure}

In Fig.~\ref{fig:mh_algorithm} a flowchart of the MH algorithm can be found.
MH consists of an alternating sequence of MD trajectories and local geometry optimizations.
The MD part allows to cross barriers, while the local geometry optimization part can locate new minima after a successful escape from the previous catchment basin. 
The velocities in the MD part are initialized according to a Maxwell-Boltzmann distribution at temperature $T$. 
Additionally these initial velocities are preferentially aligned along the direction of the soft modes~\cite{schonborn:2009}.
The Bell-Evans-Polanyi principle~\cite{jensen:1998} states that exothermic reactions have a low activation barrier. Therefore, it is favourable to cross low energy barriers to increase the probability of finding low energy structures. To make use of this principle, the temperature $T$ is adjusted dynamically, i.e., increased if the same minimum is found again and reduced if new minima are discovered.
This ensures, that just enough kinetic energy is present to overcome the lowest barriers.
The length of the MD is determined by counting the number of potential energy oscillations along the trajectory and is stopped after a given number of oscillations. 
This number is usually between 2 and 10.  
The larger it is, the further away from the current configuration the hop will lead.
Afterwards a local geometry optimization is performed.
If a new minimum has been found it is checked if the minimum already exists in the database. 
If a minimum already exists in the database the temperature is increased otherwise it is decreased.
Due to this temperature feedback the simulation cannot be trapped in a funnel.

\section{Software Description}
\label{sec:Description}

\subsection{Setting up a calculation}
\subsubsection{Input structures}
Any physically reasonable structure is valid as an input structure.
It is also possible to provide multiple structure as a list of atoms object. 
The energetically lowest will then be used as a starting structure. 
If a MPI run is started it is recommended that each worker is started with a different initial structure so that the simulation starts at different points on the PES.

\subsubsection{Potential energy surface}
Our implementation of the MH algorithm can be used with a variety of different potential energy surfaces with different characteristics. Potential energies can be obtained using analytical expressions such as force fields and MLPs or iterative procedures that are typically used in most \abinit{} methods. 
The main difference between the two methods, is that the latter typically contains numerical noise while force fields and MLPs are usually as accurate as the floating point precision.\\
In order to reliably determine whether a newly found local minimum is identical to an already found local minimum, it is necessary to do geometry optimizations with relatively tight convergence criteria. 
In our calculations, we often stop geometry optimizations after all forces are smaller than $0.01\,\sfrac{\text{eV}}{\text{\AA}}$. 
It is of great importance that energies and forces are calculated accurate enough, so that the geometry optimization convergence criterion can be reached.
In a plane wave DFT code this might for instance require a somewhat higher plane wave cutoff.
For force fields and MLPs, this is typically no problem. 
In that case it may be beneficial to choose a tighter convergence criterion for geometry optimizations such that the energy is converged to many decimal places. 
That way structures can be distinguished using only the energies and there is no need to calculate relatively expensive fingerprints.

\subsubsection{Structure distinction}
To determine correctly whether newly found minima are identical or not to the previously found minima is essential for the correct working and high efficiency of MH. 
If the energy can not be converged to many decimal places, this is done by calculating fingerprint distances between pairs of structures. 
The distances are obtained from the overlap matrix fingerprint (OMFP)~\cite{sadeghi:2013, zhu:2016}. 
The threshold distance (\textit{fingerprint\_threshold}) at which two structures are considered distinct is an important parameter and is dependent on the physical properties of the system, the noise level of the method used to calculate the energy and forces as well as the settings used to do the geometry optimization. 
A correct choice of this parameter is essential for the success of a MH simulation. 
If the threshold is chosen too small, identical local minima will mistakenly be identified to be different. 
Due to the feedback mechanism the temperature of the MD escape will drop exponentially. All the local geometry optimizations will just produce slightly different copies of the same local minimum.
If the threshold is chosen too large, different minima might incorrectly be identified to be equal.
In this case the temperature of the MD escape step increases strongly and uninteresting high energy regions are sampled.
To assist the user in finding optimal threshold values, we provide a tool to estimate these values.
In a pre-processing step a few very short MD runs at low temperature are performed followed by a relaxation. To make sure that the relaxations from all the final MD configurations lead to the same local minimum, the temperature has to be low enough to prevent an escape from the initial catchment basin.
Afterwards, the OMFP distances between all relaxed minima are evaluated. 
The pre-processing can be performed using the \textit{adjust\_fp} class. 
The result of the pre-processing is a lower bound of the OMFP threshold.
\\
For the calculation the OMFP distance between two structures, an optimal assignment is performed using a modified Jonker-Volgenant algorithm~\cite{Crouse2016Transactions} as implemented in \textit{SciPy}.
Calculating all distances between a newly found minimum and all minima in the database is computationally expensive especially if a large database is present.
Since structures which have a high energy difference are different anyway it is only necessary to compare structures within the noise level of the energy and force calculation using an OMFP distance.
Therefore, a second parameter \textit{energy\_threshold} is introduced which specifies an energy window.
If the energy difference of two minima is lower than this threshold the structures are distinguished using the OMFP distance. Otherwise the structures are considered as different.

\subsection{Features}
The algorithm described above is implemented in the \textit{MinimaHopping} class and contains a variety of features.
In order to start MH an initial structure is read in and an ASE atoms object is created. 
This is the minimal input MH needs to start.
However, it is recommended to use the above described pre-processing tool to fine tune various parameters.
A full list of the parameters and their explanation can be found in our documentation.
Several output files are generated containing either information about the progress of the simulation or are necessary for restarting the algorithm.
A detailed description of the output can also be found in the documentation.
A restart can be performed by starting the previous run again in the same directory. 
It is automatically checked if restart files have already been generated and, hence, decided if MH is restarted or started from scratch.
Several features have already been implemented in several previous versions or are newly implemented in this version.
In the following sections the most important features are presented and explained.

\subsubsection{Variable Cell Shape MD and Optimization}
For the periodic case a variable cell shape MD is implemented where not only the positions but also the lattice is adjusted.
After the MD a local geometry optimization is performed using the SQNM~\cite{schaefer2015stabilized,gubler:2022} where in the periodic case also the lattice is optimized. Moving also the lattice vectors in the MD and the optimization is necessary since the exploration of the PES should not be restricted by a fixed unit cell.

\subsubsection{Fixing Fragmentation}
Even though the temperature is kept in general as low as possible it can increase fast if the simulation is trapped in funnel. 
Performing MD of free systems at a high temperature can then result in its fragmentation which is not desirable in the context of global structure prediction.
This issue is addressed by using the DBSCAN clustering algorithm~\cite{Ester:1996, schubert:2017} to check for fragmentation of the system. 
If this is the case the velocities are changed so that the velocity of each fragment points to the center of mass of the system until no fragmentation is observed, instead of progressing with the MD.

\subsubsection{Temperature feedback}
The temperature is adjusted dynamically in the minima hopping algorithm. 
If a minimum has already been found the temperature is increased.
This can be done in two ways: (i) increasing the temperature by a fixed factor $\beta_{O}$ or (ii) by a so called enhanced feedback mechanism~\cite{schonborn:2009} where the adjustment of temperature depends on the number of times a minimum has already been visited:
\begin{equation}
    T = \beta_{O} (1 + \ln(n_{visits})) T.
\end{equation}
This enhanced feedback is recommended for very large and deep funnels where MH with the standard feed-back would have to visit a large number of minima before it is able to escape from the funnel.
As a rule of thumb, enhanced feed-back might be worth trying out if many minima are visited more than 10 times in a MH run with the standard feedback. For systems with small funnels the enhanced feed-back can lead to a less thorough search of low energy minima and, therefore, to a slight performance degradation.

\subsubsection{Construction of Datasets}
To construct high quality MLPs a dataset covering a wide range of energies and structural motives is of great importance.
The MD part of MH can provide exactly this kind of data.  
Actually it provides these data for regions around many different local minima and also for the regions around the transition states. 
So it is sampling the physically important regions of configurational space.
This makes the MH method very suitable for collecting reference data for machine learning applications. 
The flag \textit{collect\_md\_data} can be set to \textit{True} for collecting the MD data.

\subsubsection{MPI parallelization}
Best parallel efficiency of MH is achieved by running multiple runs simultaneously that access a common database containing the history of previously visited minima. 
Consequently an MPI parallelization scheme is implemented in this work where the main process is managing the database and the workers are performing MH runs.
Since multiple slave processes are sharing the same database, the temperature feedback of one MH run is influenced by all other processes.
In order to demonstrate the efficiency of the MPI parallelization a search for the ground state of the 75 atoms Lennard-Jones cluster (\ch{LJ75}) was performed. 
The global optimization of this system is particularly difficult since it is a double funnel system and the global minimum is located in a very narrow funnel~\cite{doye:1995}.
A statistical analysis was performed over 10 runs with 10 workers each, which were communicating in one case but not in the other. 
The initial structures were generated by cutting a sphere out of a cubic lattice and removing randomly atoms from the outermost and second outermost shell. 
The MPI communication improves the efficiency of the global optimization by approximately 25\% visiting on average 15000 distinct minima compared to 19000 without communication.
The most difficult initial structure is the second lowest local minimum, which is the lowest energy icosahedral structure. 
In this case all MH runs have to escape from this large funnel to find the fcc like ground state. This choice gives thus essentially a measure of how fast the algorithm can escape from the icosahedral funnel. 
On average over 10 runs, 27000 distinct minima were visited before the global minimum was found compared to 57000 without MPI communication speeding up the simulation more than 50\%.
Furthermore, the MPI parallelization is implemented such that a two level parallelization is possible by using so called group communicators.
Multiple MH processes run in parallel each of which can consist of multiple MPI processes that can be used to parallelize the energy and force evaluation.
Hence, it is possible to run several MH processes which on the other hand run the energy and force evaluation also on several MPI processes.

\subsubsection{Second Calculator}
In some cases having a second, numerically cheaper,  calculator for the MD part and a pre-optimization is useful. 
This can be done by initializing two calculators one attached to the atoms object and the other given as an argument to the \textit{MinimaHopping} class.
The MD and a pre-optimization are performed using the calculator given to the \textit{MinimaHopping} class.
The actual geometry optimization is performed using the calculator attached to the atoms object.
For the second calculator, Density Functional Theory (DFT) accuracy settings could for instance be looser, since the MD part may not necessarily require highly precise energy and force calculations. 
This flexibility can save computational resources.
A further example are biased runs where the second calculator includes a bias whereas the geometry optimization occurs on the PES without bias. Another notable application is the combination of MLPs and DFT where the MD is performed using the MLP.

\subsubsection{Reaction Network Graphs}
MH finds new structures by looking for reaction/transformation pathways with low activation energies. 
These pathways are of great interest when reactions/transformations of materials are studied. 
Our implementation keeps track of all the pathways that are found and stores them in a directed graph. 
For the graph operations the NetworkX~\cite{networkx} Python package is used and for plotting the graphs PyGraphyz~\cite{pygraphviz}.
The nodes in our graph structure are local minima and the edges are pathways connecting the local minima. 
The graph is directional and the weights of the edges are the energy difference between the current local minimum and the highest energy on the pathway.
The pathway consists of a short MD simulation followed by a geometry optimizations. 
A selection of structures that are visited on the pathway are also stored in the edges of the graph.
In order to visualize the graph, a command line tool is provided that can plot the graph, export the graph to the standard dot file format and simplify the graph. The graph can be simplified using two methods. 
Firstly, nodes with only one edge can be removed. When this is done, the parent node will be drawn slightly bigger for each child node with only one edge that is removed. 
Secondly, multiple nodes in a row with two edges can be contracted into a single edge.

\subsection{Analyzing and Post-processing}
Beside the generation of reaction network graphs some command line tools are implemented assisting the post-processing:
\begin{itemize}
    \item \textit{omfpdistance}: Gives the OMFP distance of two structures.
    \item \textit{sortAtoms}: Sorts structures according to their energy. 
    \item \textit{standardizeLattice}: Uses the \textit{spglib} to standardize the lattice vectors.
\end{itemize}
However, further post-processing is highly dependent on the system and the research task which is to be solved.
For example, a saddle point search can be performed connecting the minima found by MH to construct reaction pathways~\cite{schaefer:2014,de:2019, de:2020}.
In another study the MH algorithm was used to discover different polymorphs of MaPbI~\cite{flores:2018} which were then used as input structures for the Funnel Hopping Monte Carlo method~\cite{finkler2020funnel, finkler:2023} to identify the free energy of the different phases.

\section{Illustrative Example}
\label{sec:Example}
\begin{figure}
    \centering
    \includegraphics[width=0.5\textwidth]{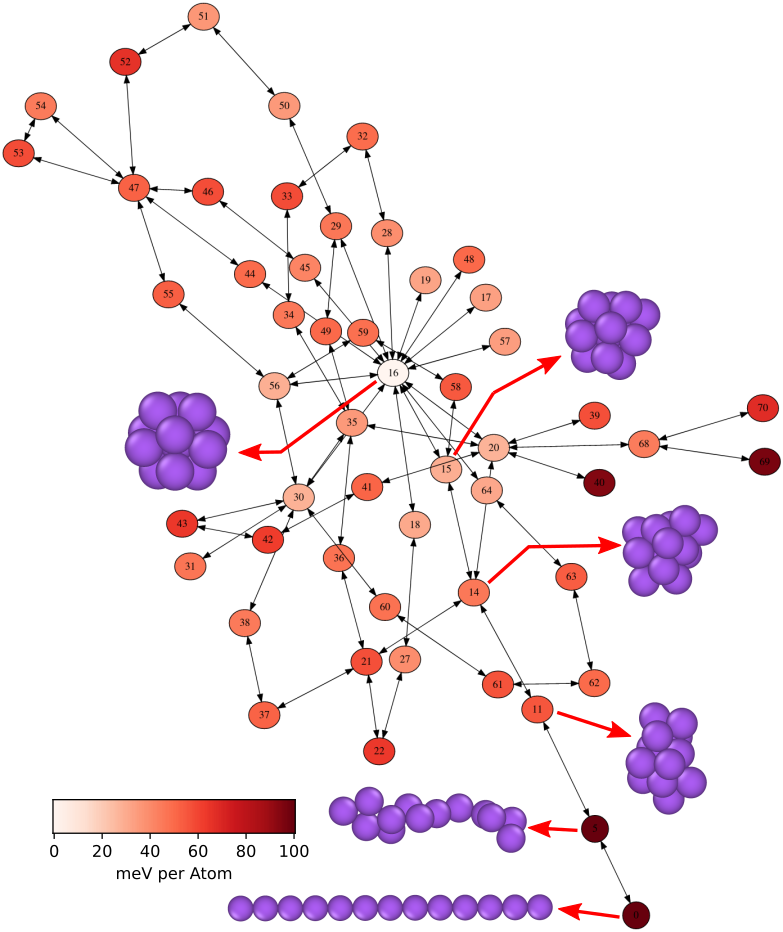}
    \caption{Graph of one MH run with an \ch{Na13} cluster starting with an chain (0) to the global minimum (16). Each node indicates a minimum and the color of the nodes are displaying its energy difference to the global minimum (16). Also the structures along the shortest pathway from the initial structure to the global minimum are displayed.}
    \label{fig:graph}
\end{figure}

In this example the global optimum of an \ch{Na13} cluster is determined. 
In the initial configuration the atoms are arranged as a chain. 
Even though this initial guess is very far from the global minimum the MH algorithm is able to find the ground state very fast.
The underlying energy and force evaluation is performed using an EAM force field~\cite{daw:1983} implemented in ASE and parameterized for sodium~\cite{nichol:property}.
The global minimum is an icosahedral structure which has also been found using other potentials~\cite{noya:2007}.
\\
In the pre-processing a lower bound of the OMFP threshold of \textit{fingerprint\_threshold}$ = 2.4 \cdot 10^{-4}$ was determined.
\\
The MH run is performed using the \textit{examples/mh\_na13.py} script.
First, the initial structure is generated using ASE.
Afterwards, the EAM calculator is set up.
Next, the MH class is built using the same parameters as in the pre-processing.
For the MH run a threshold of \textit{fingerprint\_threshold}$ = 10^{-3}$ is chosen for the distinction of different minima.
Files sorted by energy of the lowest minima can be found in the \textit{minima} folder. 
Since the ground state in this example is known, it can be checked visually if the icosahedral structure has been found.
If the ground state is not known an often applied approach is stop the run if the same lowest energy structure has been found multiple times.
Moreover, a restart folder is present in order to restart the MH algorithm.
Using the default settings of the MH algrithm an average of 22.5 MD--optimization cycles are needed to find the icosahedral ground state of \ch{Na13}.
\\
If we evaluate the run we can create a graph with the command line interface \textit{parseGraph}, see Fig.~\ref{fig:graph} , which shows the local minima found as nodes that are connected to each other if the minima hopping algorithm directly hopped from one local minimum to another. 
Each minimum is labeled with a unique, positive integer that is also displayed in the graph.
Furthermore, it can be seen in the graph that the simulation does not stop once the global minimum has been found.
In the depiction of the graph presented here also snapshots of the structures along the shortest path towards the global minimum are shown.

\section{Conclusion}
This paper presents a Python implementation of the Minima Hopping algorithm.
The installation and usage of the software are both straightforward and user-friendly.
With the new implementation of the MH algorithm it is possible to explore the PES of materials very efficiently using the implemented MPI parallelization scheme where multiple minima hopping processes share one single database.
Notably, this approach has demonstrated significant accelerations in global optimization, particularly for complex PES such as the LJ75 cluster, where the number of energy and force evaluation was cut in half compared to the serial minima hopping implementation. 
The new implementation is highly flexible in terms of systems and calculators. 
Hence, interfacing minima hopping to the ASE library makes a broad variety of calculators accessible based on DFT, tight-binding, force fields and MLPs. 
For the simulation bulk systems, variable cell shape softening, MD and geometry optimization are implemented so that the exploration of the PES is not limited to a particular lattice.
The combination of MLPs and the MPI variant of our implementation makes it possible to predict ground states of large and complicated systems rapidly with \abinit{} accuracy.


\appendix


 \bibliographystyle{elsarticle-num} 
 \bibliography{cas-refs}





\end{document}